\documentclass[useAMS,usenatbib, usegraphicx]{mn2e}
\usepackage{amssymb, hyperref, rotating, multirow}
  \def\apjs{ApJS} 
\def\apjl{ApJ}   \def\aap{A\&A}
   
\let\ga=\gtrsim \let\la=\lesssim
\def\kms{km\thinspace s$^{-1}$ }     

\title[BH growth in MS galaxies] 
{Simulations of supermassive black hole growth in high-redshift disk galaxies}

\author[Gabor et al.]{
J. M. Gabor,$^{1}$\thanks{Email:jared.gabor@cea.fr}
Fr\'ed\'eric Bournaud$^{1}$ 
\\ $^{1}$CEA-Saclay, IRFU, SAp, F-91191 Gif-sur-Yvette, France  
}

\begin{document}


\pagerange{\pageref{firstpage}--\pageref{lastpage}} \pubyear{2013}

\maketitle
\label{firstpage}

\begin{abstract}
Observations suggest that a large fraction of black hole growth occurs
in normal star-forming disk galaxies.  Here we describe simulations of
black hole accretion in isolated disk galaxies with sufficient
resolution ($\sim 5$ pc) to track the formation of giant molecular
clouds that feed the black hole.  Black holes in $z\sim 2$ gas-rich
disks ($f_{\rm gas}\approx 50$\%) occasionally undergo
$\sim10$~Myr episodes of Eddington-limited accretion driven by
stochastic collisions with massive, dense clouds.  We predict that
these gas-rich disks host weak AGNs $\sim1/4$ of the time, and
moderate/strong AGNs $\sim 10$\% of the time.  Averaged over
$\sim$100~Myr timescales and the full distribution of accretion rates,
the black holes grow at a few per cent of the Eddington limit --
sufficient to match observations and keep the galaxies on the $M_{\rm
  BH}-M_{\rm bulge}$ relation.  This suggests that dense cloud
accretion in isolated $z\approx2$ disks could dominate cosmic black
hole growth.  In $z\sim 0$ disks with $f_{\rm gas}\approx 10$\%,
Eddington-limited growth is extremely rare because typical gas clouds
are smaller and more susceptible to disruption by AGN
feedback.  This results in an average black hole growth rate in
high-$f_{\rm gas}$ galaxies that is up to $10^3$ times higher than
that in low-$f_{\rm gas}$ galaxies.  In all our simulations, accretion
shows variability by factors of $\sim10^4$ on a variety of time
scales, with variability at $\sim 1$~Myr scales driven by the
structure of the interstellar medium.
\end{abstract}


\section{Introduction} 
Supermassive black holes (BHs) are widely thought to be linked to galaxy
evolution.  Correlations between black hole mass and galaxy properties
suggest the imprint of co-evolution \citep[e.g][]{richstone98,
  ferrarese00,gebhardt00,marconi03,haring04}.  Active Galactic Nuclei
(AGN), the energetic signposts of black hole growth, provide a
physical link between black hole growth and the wider galaxy
\citep[e.g.][]{silk98,dimatteo05}.  Yet the physical processes of black hole
fueling that trigger AGN are not fully understood.

At first glance, moving gas from galactic scales to the BH poses a
huge angular momentum problem \citep[cf.][]{jogee06}.
One effective way to generate sufficient torques
is via major, gas-rich mergers between galaxies.  Simulations show
that major mergers funnel gas toward the central black holes
\citep[][]{hernquist89, barnes92, dimatteo05}.  Owing to the
effectiveness of this merger-triggering mechanism, and because the
global torques are relatively easy to resolve, a great deal of
computational and theoretical work has investigated various aspects of
merger-induced BH growth
\citep[e.g.][]{springel05_mergers_ellipticals, hopkins05,
  hopkins08_ellipticals,johansson09_BH,debuhr10,debuhr11}.

Although mergers probably do enhance BH activity \citep{ellison11},
observations have not strongly supported the major merger scenario for
most AGNs.  AGNs typically live in ``normal,'' isolated galaxies, with
no obvious signs of a major merger \citep{grogin05, pierce07, gabor09,
  georgakakis09, cisternas11, kocevski12}.  While some observations
indicate that AGN host galaxies tend to be bulge-dominated
\citep[e.g.][]{kauffmann03_agn_hosts, grogin05}, such hosts are in the
minority at higher-redshifts and AGN are common in star-forming disks
\citep{gabor09, hwang10, cisternas11, kocevski12, schawinski12,
  juneau13}.  Black hole growth, like star-formation, mostly seems to
occur in the Main Sequence of star-forming disk galaxies
\citep{mullaney12, rosario12}.

These observations suggest that an important alternative mechanism
must drive black hole growth in isolated, gaseous disk galaxies.
\citet{hopkins_hernquist06} suggest that interstellar clouds in disk
galaxies, owing to velocity perturbations away from a purely Keplerian
orbit, will randomly scatter close enough to fuel the central BH.  In
very gas-rich galaxies such as those observed at high-redshift
\citep[e.g.][]{daddi10, tacconi10}, instabilities form clumps and
transient features, and interactions among these perturbations cause
torques that drive a global inflow \citep{dekel09, ceverino10,
  bournaud11}.  This inflow, which does not rely on the clumps being
long-lived, drives central (bulge) star-formation and could feed the
black hole.  Whether this instability-driven inflow can remove enough
angular momentum to reach the central black hole is not clear, but
only a small fraction of the global mass inflow needs to reach the
black hole to drive substantial growth.

In this work we present high-resolution hydrodynamic simulations of
isolated disk galaxies with a model for black hole fueling and
feedback.  With a spatial resolution of $6$ pc, these simulations can
resolve the disk scale height and individual clouds in the ISM of the
galaxies.  This allows us to capture angular momentum exchange among
gas clouds that can drive gas toward the galactic center.  We focus on
two broad regimes of disk galaxies -- high-redshift clumpy disks with
$\sim50$\% gas fractions, and low-redshift disks with $\sim$10\% gas
fractions -- and our simulations span a range of stellar masses.


As we show below, massive, dense gas clouds in high-redshift gas-rich
galaxies stochastically collide with the black hole and drive
persistent episodes of Eddington-limited accretion.  The accretion
rates are sufficient to explain most of the black hole growth at
$z\sim2$.  This contrasts starkly with BH growth in low-$z$, low-gas-fraction
galaxies, which is tiny and never reaches the Eddington limit.  A
difference in gas fraction of $\times5$ leads to a difference in BH
growth of up to $\times10^3$.

We describe the simulations in \S \ref{sec.sims}.  In \S
\ref{sec.results} we describe the basic results of the simulations,
then in \S \ref{sec.implications} we extend the results to trends in
global cosmic black hole growth.

\section{Simulations} 
\label{sec.sims} 
\begin{figure*}
\includegraphics[width=168mm]{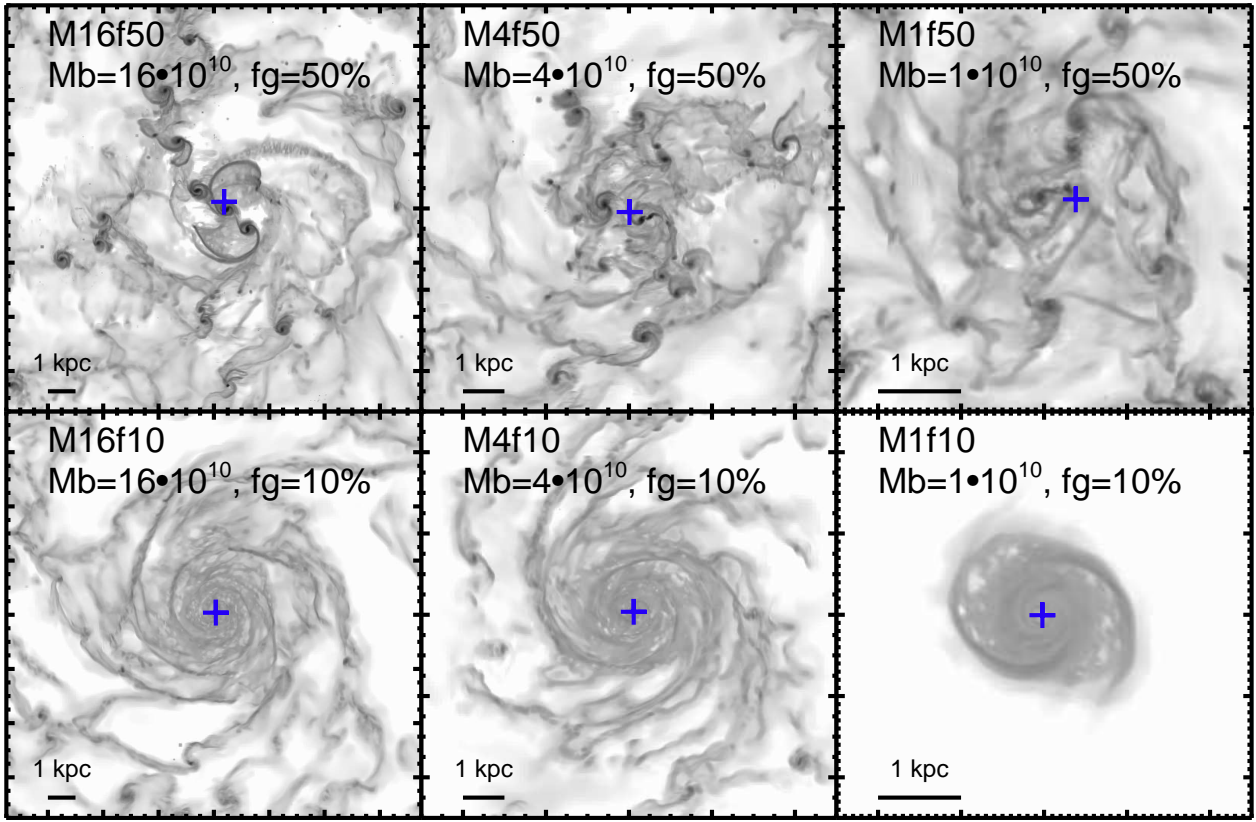}
\caption{Face-on projected images of mass-weighted gas density for
  each simulation, as labelled.  A blue cross shows the position of
  the black hole particle.  Gas-rich $z=2$ galaxies (top row) form dense
  clouds and giant clumps that migrate through the disk, while
  low-gas-fraction galaxies (bottom row) form tight spiral arms.}
\label{fig.example_snapshots}
\end{figure*}
We have run a small suite of six fiducial simulations, plus some
variations in resolution and physics recipes.  Here we describe the
hydrodynamics code along with models for cooling, star-formation, and
black hole physics.  Then we describe the details of the simulation
runs, including initial conditions and relaxation.

\subsection{Hydrodynamics with Ramses}
We use the Adaptive Mesh Refinement (AMR) hydrodynamics code Ramses
\citep{teyssier02}, version
3\footnote{\url{http://www.itp.uzh.ch/\~teyssier/Site/RAMSES.html}},
and we use customized code patches to alter various aspects of the
physics.  Ramses solves the equations of hydrodynamics and gravitational
N-body dynamics on a Cartesian grid whose cells can vary in size.
Dark matter and stars are treated as particles and behave as
collisionless fluids.  The Poisson equation is solved for each cell of
the mesh, and the acceleration of each particle is then found by
interpolating the potential to the position of the particle within the
cell.

A base grid is set up at a given level of refinement with a fixed
number of cells per side.  For refinement level $n=8$ (which is the
coarsest level in our simulations), there are $2^n = 256$ cells per
side $\rightarrow 256^3 \sim 10^7$ total cells.  Each of these cells
may be \emph{refined} based on some physical criteria.  We refine
cells when the total mass in a cell exceeds $6\times10^5 M_{\rm sun}$,
the number of dark matter particles in a cell exceeds 30, or the local
gas Jeans length in a cell is not resolved by at least 4 cell widths
\citep{truelove97}.  When refined, a cell is divided equally into 8
smaller cubic cells, thus giving a higher spatial resolution (by
$\times 2 $).  These smaller cells may also be refined into smaller
and smaller cells, up to some smallest-cell limit determined by the
user.  We use a maximum level of refinement of $n=13$, so that in our
50~kpc simulation box the maximum resolution is 50~kpc$/2^{13} = 6$~pc
(corresponding to a maximal mass resolution of $\sim250M_{\rm sun}$ for
typical ISM gas).  This adaptive refinement enables high resolution
where it is required without wasting resources by applying high
resolution where it is not required.  Timesteps are adapted to the
levels of refinement so that the timestep for cells at refinement
level $n$ is twice as long as the timestep at level $n+1$.

Ramses incorporates cooling, star-formation, and supernova feedback.
Cooling acts as a sink for the gas thermal energy in each grid cell.
In our simulations, we assume all the gas has solar metallicity.  We
allow cooling down to a floor of 10~K, but we also enforce a
density-dependent temperature floor to prevent artificial fragmentation
in dense regions.  This temperature floor explicitly fulfills the
requirement that the local Jeans length is always resolved by at least
four grid cells \citep{truelove97}.

Star-formation occurs in gas cells of sufficiently high density.  We
use a density threshold of 100 H atoms cm$^{-3}$.  At
each timestep $dt$, the star-formation rate (SFR) in a gas cell above the
threshold is 
$\dot{M}_* = 0.01 V_{\rm cell} \rho_{\rm gas} / t_{\rm ff}$, 
where 0.01 is the star-formation efficiency
  \citep[cf.][]{zuckerman74, krumholz07}, $V_{\rm cell}$ is the cell
  volume, $\rho_{\rm gas}$ is the gas density, and $t_{\rm
    ff}=(3\pi/32G\rho_{\rm gas})^{1/2}$ is the free-fall time.  Then,
  new star particles of mass $\approx5.8\times10^3 M_{\sun}$ are
  spawned with a probability consistent with the SFR $\dot{M}_*$ and
  timestep $dt$.  Once a particle is spawned, its mass is subtracted
  from the gas cell.  If the desired star particle mass exceeds half
  the gas mass in the cell, then no star particle is allowed to form.
  This prevents density discontinuities due to the sudden drop in
  density.  Such star-formation ``lost'' due to the mass criterion is
  reported by the code, and is typically a few percent.  When a star
  particle does form, it inherits the position and velocity of its
  parent cell, but then evolves independently from then on.

Although we have tested various models for stellar feedback, here we
use a simple thermal model for supernova.  In this model, a star
particle dumps thermal energy into its host grid cell's gas 10~Myrs
after it is spawned (its host cell at $t_{\rm age}=10$~Myrs may differ
from the cell that originally spawned it at $t_{\rm age}=0$).  The
star particle represents a star cluster 10\% of whose initial mass
explodes as supernovae.  For each $10M_{\sun}$ of progenitor mass
which explodes as a supernova, $10^{51}$~ergs of energy is
instantaneously added as thermal energy to the nearest gas cell.  We
delay cooling for this heated cell \citep[cf.][]{stinson06} for 2~Myrs
to prevent the injected energy from being rapidly radiated away.  This
cooling delay tends to suppress the overall SFR of the galaxy (and
bring it roughly into agreement with observations) and allows heated
gas to gain substantial momentum and generate winds.

\subsection{Black hole fueling and feedback model}
The publicly available version of Ramses also incorporates a model for
the formation, fueling, and feedback of supermassive black holes.
Because this model was originally implemented for cosmological
simulations, we have made several alterations to make it more
appropriate for idealized, high-resolution simulations of individual
galaxies.  The black hole is represented by a collisionless sink
particle, much like the dark matter and star particles, which is added
to the initial conditions.  Black hole accretion is handled with a
Bondi formalism based on \citet{krumholz04}, and AGN feedback is
implemented with a model similar to that described in
\citet{teyssier11} and \citet{dubois12_dual}, which is partly based on the work of
\citet{booth09}.

\subsubsection{Black hole accretion}

The accretion rate on to the black hole is given by a Bondi \citep{bondi44, bondi52} formula of
the form: 
\begin{equation}
\dot{M}_{\rm BH} = \alpha \frac{4 \pi G^2 M_{\rm BH}^2 \rho}{(c_s^2 + u^2)^{3/2}}.
\end{equation}
$M_{\rm BH}$ is the BH mass, $G$ is the gravitational constant, $\rho$
is the local gas density, $c_s$ is the local gas sound speed, and $u$
is the velocity of the gas relative to the BH.  Properties of the
surrounding gas are measured within an accretion region of radius
$r_{\rm acc} = 4\Delta x \approx 24$~pc, where $\Delta x$ is the smallest resolution
element of the simulation.  The gas properties are computed as
weighted averages within the accretion region, with gas elements close
to the black hole having larger weights \citep{krumholz04}.  We have
chosen $\alpha = 1$, so there is never a ``boost'' factor in the
accretion rate.

The black hole accretion rate is capped at the Eddington limit given by
\begin{equation}
\dot{M}_{\rm Edd} = \frac{4 \pi G M_{\rm BH} m_p}{\epsilon_r \sigma_T c},
\end{equation}
where $m_p$ is the proton mass, $\sigma_T$ is the Thomson scattering
cross-section, $c$ is the speed of light, and $\epsilon_r$ is the
efficiency with which accreted mass is converted into luminous energy.
We use $\epsilon_r = 0.1$.

At each timestep (of duration $dt$) a gas mass of $\dot{M}_{\rm
  acc} dt = {\rm min}\{ \dot{M}_{\rm BH}, \dot{M}_{\rm Edd} \} dt$ is
  removed from the surrounding gas.  Gas is removed from cells within
  $r_{\rm acc}$ using the same weighting scheme as for averaging gas
  properties.  In order to prevent zero or negative gas densities, the
  fraction of gas removed from a given cell is limited to 50\% of its
  original gas mass.  The actual amount of mass accreted is then
  stored as a variable, since it will be used at coarse timesteps to
  determine the amount of AGN feedback.

\subsubsection{AGN feedback}

At each coarse timestep (the timestep corresponding to
the coarsest level of refinement), we inject thermal energy into the
gas surrounding the black hole.  The injected energy is given by
\begin{equation}
\Delta E_{\rm acc} = \epsilon_c \epsilon_r \dot{M}_{\rm acc} c^2 dt
\end{equation}
Here $\epsilon_c$ is the efficiency with which radiation from
accretion couples to matter.  We adopt the fiducial value of
$\epsilon_c=0.15$ used by \citet{booth09} and \citet{teyssier11}, and
$\epsilon_r=0.1$ as above.

The standard implementation of AGN feedback attempts to inject this
energy at uniform pressure throughout the accretion region.  If the
available thermal energy is insufficient to heat the gas to a weighted
average temperature $T_{\rm min}=10^7$~K, then no injection occurs and
the accretion energy $\Delta E_{\rm acc}$ is stored and added to that
in the following timestep.  This minimum temperature ensures that once
an AGN outburst occurs, its energy will not be radiated away so
rapidly as to have no effect on the gas.  If the available thermal
energy is sufficient to heat the gas above a maximal temperature
$T_{\rm max}=5\times10^9$~K, then only enough energy is injected to
heat the gas to that average temperature.  In this case, the unused
energy is again stored for use in the next timestep. This maximum
temperature is probably implemented for technical reasons to prevent
superheating of the gas, and correspondingly short timesteps which can
grind a simulation to a halt (although it fails to prevent this
problem, as described in the following paragraph).  The maximum
temperature is problematic in some cases, however.  If the BH is in
the midst of a very dense clump of gas, its Bondi rate may exceed the
Eddington limit for a very long time.  The energetic AGN output is not
enough to blow away the dense clump, and the unused reserve of energy
builds up to very large values as it is carried over from timestep to
timestep.

Due to the clumpy ISM structure in our high-resolution
simulations, a uniform-pressure injection sometimes yields
unphysically high thermal temperatures ($>10^{15}$~K) -- since
pressure $P \propto \rho T$, injecting a huge amount of pressure into
low density gas gives a large temperature.  As a practical matter,
such large temperatures lead to very small timesteps (thanks to the
Courant condition), which make the simulations very expensive or
impossible to run.

We have changed the AGN feedback implementation to circumvent these
problems as follows.  We inject the AGN outburst only into cold, dense
gas instead of injecting a uniform pressure.  We use an iterative
procedure.  First, we do a trial energy injection in the normal way,
and check the post-injection temperature of all gas cells.  Any cell
whose post-injection temperature is above $T_{\rm max}$ is marked to
be excluded from energy injection.  If any such cells exist, then we
iterate by applying a trial energy injection to all cells except the
excluded ones.  We continue this process until all cells'
post-injection temperatures are below $T_{\rm max}$.  This procedure
is only rarely necessary.  When we do execute it, $<10$\% of all
available cells are excluded, and the typical density of these cells
is $\sim0.1$~H~cm$^{-3}$, thus representing a tiny fraction of the mass
within the blast region.

In cases where the post-injection temperature would exceed $T_{\rm
  max}$ no matter how the outburst energy is distributed, we expand
the blast region.  We multiply the blast radius by factors of 1.25 (to
roughly double the volume) in an iterative procedure.  After
increasing the blast region, we re-run the injection scheme and check
whether all the outburst energy can be used without the temperature
exceeding $T_{\rm max}$.  If not, then we expand the blast radius
again, and iterate.  In the simulations presented here, this change
rarely makes a significant difference because long periods of
Eddington-limited accretion are rare.

\subsubsection{Black hole's motion}

For the purposes of calculating the motion of the black hole, the mass
of the black hole is evenly distributed over a few thousand particles
within the accretion region.  These particles are re-generated in a
uniform spherical cloud at every coarse timestep.  Between coarse
timesteps (at finer timesteps), the motion of each collisionless
``cloud'' particle is tracked separately, and the black hole always
inherits the velocity of its nearest cloud particle.  When black hole
accretion occurs, the momentum of accreted material is added to the
black hole.

In test simulations at high resolution we found that this default
implementation of black hole motion led to large scatter of the BH
particle from the center of the galaxy, up to several kpc within the
disk.  In gas-rich disk simulations the local center-of-mass at the
center of a galaxy (usually dominated by bulge stars) does indeed move
out of the exact center, but by a few hundred pc, not several kpc.
The scattering appears to be driven by interactions with massive,
dense, star-forming clouds, but it is not clear whether this large
black hole scattering is realistic or if it depends on the details of
our implementation.  To avoid this difficulty, we implemented an
alternate formulation where the black hole tracks the local center of
stellar mass.  At each timestep we calculate the center of mass of
stars within 100~pc of the BH (which in our simulation is well outside
the sphere of influence of the BH). Instead of explicitly resetting
the position of the BH, which could lead to discontinuous jumps in BH
position, we instead set the BH velocity towards that center of mass.
The magnitude of the velocity is $0.5 S_{\rm COM}$~\kms~pc$^{-1} +
v_{\rm COM}$, where $S_{\rm COM}$ is the separation between the BH and
the center-of-mass, and $v_{\rm COM}$ is the velocity of the
center-of-mass.  In this formulation the momentum of accreted gas is
ignored.

\subsubsection{Other black hole models}
 Although our Bondi accretion plus thermal feedback model is similar to
those most commonly used, several alternatives have emerged in recent
literature.  For accretion, alternative models include those based on
viscous disk accretion \citep{debuhr11}, drag force generated by
stellar feedback \citep{okamoto08}, gravitational torques
\citep{hopkins11}, accretion of low angular momentum gas onto an
accretion disk particle \citep{power11}, or the free fall time of
surrounding gas \citep{hobbs12}.  Generally, these models better
account for the angular momentum of potentially accreting gas than
does the Bondi model.  For AGN feedback, alternative models include
kinetic winds emerging from the smallest scales \citep{novak11,
  power11}, radiation-pressure driven winds \citep{nayakshin09,
  nayakshin10, debuhr11}, and variations in the details of energy
input (temperature, extent, etc.).  Although our model is meant to
represent thermal heating by AGN radiation, the enforcement of a
minimum injection temperature effectively ensures that a wind will
develop during periods of rapid accretion.

Given the variety of possible model combinations, a detailed study is
beyond the scope of the present paper.  Recent comparative studies
imply that the particular choice of model (and its parameters) will
affect details of BH accretion histories, and can change the long-term
dynamics \citep{newton13, wurster13, wurster13_compare}.  Regardless
of accretion model, our simulations robustly predict the close
approach of gas within $\sim 10$~pc of the black hole, which drives
our main results.  As we will show, periods of high accretion rates
correspond closely to the amount of gas within $\approx 50$pc of the
black hole.  What happens when the gas gets within a few tens of pc is
model-dependent, though our results should be qualitatively
insensitive to the details.
\subsection{Simulation Runs}

\begin{table}
\caption{Simulation Runs}
\begin{tabular}{rlccc}
\hline
\hline
& Simulation & $M_{\rm baryon} (M_{\sun})$ & $f_{\rm gas}(t=0)$ & $M_{\rm BH}(t=0) (M_{\sun})$ \\
\hline
\multirow{3}{*}{\begin{sideways}$z=2$\end{sideways}}&M16f50 &  $1.6\times10^{11}$  & 0.60 & $7.9\times10^7$  \\    
&M4f50  &  $4\times10^{10}$    & 0.54 & $2.4\times10^7$  \\    
&M1f50  &  $1\times10^{10}$    & 0.57 & $6.3\times10^6$  \\    
\hline
\multirow{3}{*}{\begin{sideways}$z=0$\end{sideways}}&M16f10 &  $1.6\times10^{11}$  & 0.10 & $4.3\times10^7$  \\    
&M4f10  &  $4\times10^{10}$    & 0.10 & $3.1\times10^7$  \\    
&M1f10  &  $1\times10^{10}$    & 0.10 & $5.1\times10^6$  \\    
\hline
\label{table.runs}
\end{tabular}
\end{table}

In order to bracket the properties of Main Sequence disk galaxies from
high to low redshift \citep{daddi07, noeske07, elbaz11}, we focus on three regimes in baryonic mass
($M_{\rm baryon}=1\times10^{10}, 4\times10^{10},$ and $16\times10^{10}
M_{\sun}$), and two regimes of gas fraction ($f_{\rm gas}\approx50$\%
and 10\%), for a total of six fiducial simulations (Table
\ref{table.runs}).  The galaxies with $f_{\rm gas}\approx 50$\%
represent gas-rich disks observed at $z\sim2$
(e.g. \citealt{daddi10_gasfrac, tacconi10, tacconi13, magdis12}, although see
\citealt{narayanan12}), while the $f_{\rm gas}\approx 10$\% galaxies
represent disk galaxies at $z=0$ \citep[e.g.][]{blanton09}.  We have
named the simulations by encoding their baryonic masses and
approximate gas fractions: e.g. galaxy M16f50 is the galaxy with a
baryonic mass of $16\times10^{10} M_{\sun}$ and gas fraction $\approx
50$\%.  In the future we hope to extend this small suite of
simulations to a broader range of properties.

We create initial conditions using particles for stars and dark
matter, and density on a grid for the gas disk. The particle
phase-space distribution is set up using the code from
\citep{bournaud02}, in which the gas disk density distribution is
added analytically. The rotation velocity for gas is computed from the
circular velocity of the total gravitational potential, and corrected
for the additional pressure term resulting from the density gradient
\citep[e.g. equation 13 in][]{wang10}, and the equivalent ``asymmetric
drift'' correction is applied for stars. To improve the relaxation of
the initial mass distribution, we compute the total gravitational
potential once, let the particles evolve in this frozen potential for
two disk dynamical times, and repeat this relaxation phase three times
in total. The gas density profile is frozen so that the surface
density does not vary from the initially chosen value. This
ensures that the total mass distribution is close to
equilibrium.  This technique was chosen to keep the gas surface density as
close as possible to its initial value as this is the most relevant
parameter quantifying the reservoir available for AGN feeding, even if
the dark matter density profile may be somewhat more affected in the
initial particle relaxation phase.


Each of our simulations' initial conditions includes $1.5-2$ million
total collisionless particles, divided roughly equally between dark
matter and star particles.  Since we are focused on the central parts
of the galaxy, we do not model the full dark matter halo outside the
disk. The gaseous disks have exponential radial profiles with scale
radii $2.5-10$~kpc (depending on the baryonic mass) and truncation
radii about $1.4$ times larger.  The gas has an exponential scale
height of $400 - 1000$~pc.  Both gas and stars are initially rotating
with a flat rotation curve.  Outside the disk (which represents
intergalactic space), the AMR gas cells are initialized with a uniform
density $5\times10^{-4}$ times that at the edge of the disk,
corresponding to $\sim 10^{-5}$~H~cm$^{-3}$.  We insert a black hole
sink particle in the center of the galaxy with a mass within the
scatter allowed by the local $M_{\rm BH} - M_{\rm bulge}$ relation
\citep[e.g][]{marconi03, bennert11}.

From the initial conditions, we run the simulations at low-resolution
(typically $\sim 50$~pc) for a relaxation phase of
$100-300$~Myrs. We do this because initial conditions
  incorporating a disk and spherical halo cannot be fully in
  equilibrium, and finite resolution and initial AGN feedback may cause additional perturbations.  During
  the first dynamical time, ring-shaped density waves move outward in
  the disk.  These density perturbations generally show no more than a
  factor of 3 contrast with the smooth density profile, so they do not
  trigger instabilities or convey substantial gas mass outwards.  The
perturbations are damped out after the first dynamical time, and we
allow structure such as spiral arms to develop before increasing to
the full resolution.  During this initial low-resolution relaxation
phase, we allow star-formation, black hole accretion, and the
associated feedback to prevent large gas densities from building up
rapidly.  We found that star-formation can rapidly deplete gas-rich
galaxies of gas during this phase, so we construct the initial
conditions with a gas fraction of 63\%.

After the initial relaxation phase, we switch the simulation to the
full resolution of 6.1 pc, allowing us to resolve dense structures up
to several $\times10^5$~H~cm$^{-3}$.  We denote the beginning of the
high-resolution phase as time $t=0$.  We run each simulation for
$\sim$100~Myrs, roughly a galaxy dynamical time, and several local
dynamical times in the galactic centers.  This time is sufficient to
capture the dynamical evolution and its interaction with the black
hole, but short enough that our simulations should be insensitive to
uncertainties in the lifetimes of massive clumps.  Estimates of clump
lifetimes in $z\approx2$ galaxies are $\gtrsim10^8$~yrs, even in the presence of strong
stellar feedback \citep[e.g.][]{genel12, wuyts12, dekel13}.  Over much
longer timescales (several $\times10^8$~yrs), cosmological inflows and
mergers are expected to impact the evolution of the galaxy and its
black hole \citep{dekel09_nature, dave12}, so cosmological simulations
might be a more appropriate tool \citep[e.g.][DeGraf et al. in
  preparation]{kim11, dubois12_feeding}.

In addition to the six fiducial runs shown in Table \ref{table.runs},
we have also run several variant simulations.  To test resolution
effects, we have run several models at a resolution of 12~pc, as well
as a brief periods at 3~pc resolution.  These resolution tests show
only marginal differences that do not impact our results.  We also
test the impact of AGN feedback by running two simulations -- M4f50
and M4f10 -- with all the same physics except our AGN feedback model
(\S \ref{sec.feedback_impact}).

\section{Simulation results}
\label{sec.results}

Figure \ref{fig.example_snapshots} shows face-on images of gas density
for each of our six simulations.  This figure highlights the
qualitative difference between gas-rich galaxies like those found at
$z=2$, and low-gas-fraction disk galaxies found in the local universe:
gas rich galaxies preferentially form dense clouds and clumps, while
low-gas-fraction galaxies form spiral arms and less-dense clouds.  As
described in \citet{bournaud11}, gas density perturbations such as
clumps will interact with each other, applying torques that lead to
global gas inflow toward the central regions.  This results in markedly
different black hole fueling patterns that depend on the gas fraction,
as we show below.

%
%
\subsection{Black hole accretion rates: high gas fraction vs. low gas fraction}
\begin{figure*}
\includegraphics[width=168mm]{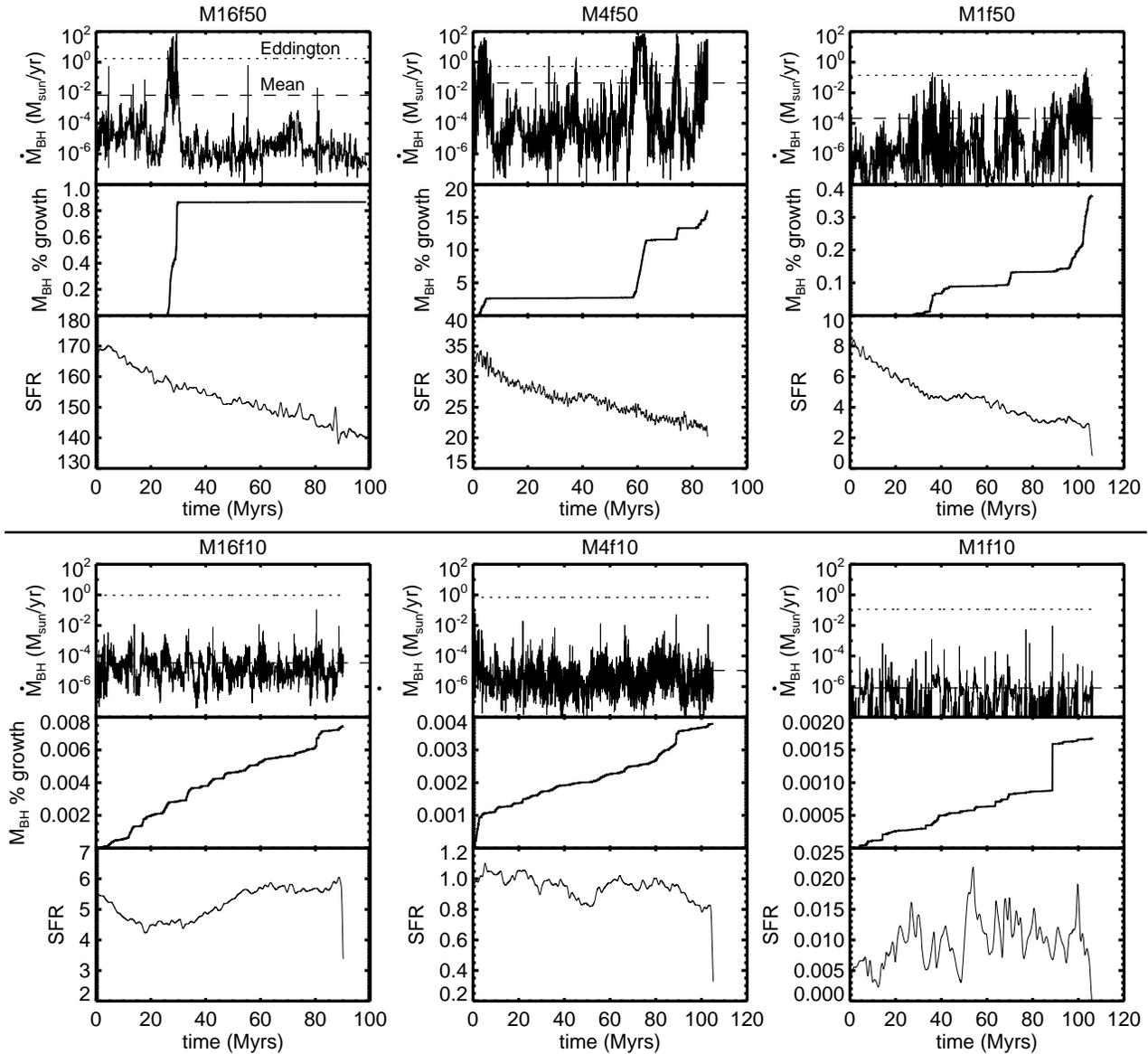}
\caption{For each of our six simulations (labelled panels), we show
  Bondi accretion rates in $M_{\sun}$~yr$^{-1}$ (top sub-panel), BH
  mass growth (middle), and star-formation rates in
  $M_{\sun}$~yr$^{-1}$ (bottom).  In the accretion rate plots, a
  horizontal dotted line shows the Eddington limit at which the
  actual BH accretion is capped, and a horizontal dashed line shows
  the mean actual accretion rate.  All simulated galaxies show highly
  variable BH accretion, but gas-rich galaxies (top row) experience
  phases of Eddington-limited accretion that dominate BH
  mass growth.}
\label{fig.time_series}
\end{figure*}

In this section, we show that black holes in our gas-rich simulations
occasionally undergo persistent episodes ($\sim 10$~Myr long) of
Eddington-limited accretion, and they grow by up to 20\% over the 100
Myr simulation interval.  Our low-gas-fraction simulations, on the
other hand, very rarely reach the Eddington limit, and grow by
$<0.01\%$, or by one part in ten thousand.

Figure \ref{fig.time_series} illustrates these results.  The top row
shows our high-gas-fraction galaxies, and the bottom row shows
low-gas-fraction galaxies.  For each galaxy, we show the Bondi
accretion rate (top panel), the per cent black hole mass growth
(middle), and the star-formation rate as a function of time (bottom).
The Bondi accretion rate is measured directly in the simulations and
reported at each coarse timestep, along with the black hole mass and
position.  We calculate the fractional BH mass growth, $g$, using the
reported BH masses and the definition $g=(M_{\rm BH}(t) - M_{\rm
  BH}(t=0)) / M_{\rm BH}(t=0)$.  We calculate the star-formation rate
by adding up the masses of star particles born during each time
interval in the simulation.

In simulations with high-$f_{\rm gas}$, the accretion rate typically
fluctuates around $10^{-5}$ of the Eddington limit (dotted horizontal
lines).  Brief fluctuations by factors $\sim 10^4$ are relatively
common, though they typically last $<<1$~Myr.  Occasionally the Bondi
rate jumps above the Eddington rate for periods of $\sim 10$~Myrs (but
the actual accretion rate in the simulations is capped at the
Eddington rate).  As shown by rapid jumps in the fractional black hole
mass growth plots, these high-accretion-rate phases are responsible
for nearly all the mass growth of the black hole.  Throughout the
simulations, the SFR evolves slowly, varying by factors of $\lesssim
2$.  Trends with baryonic mass are not apparent -- galaxy M4f50, the
intermediate mass in our suite, shows more high-accretion episodes and
thus more BH growth than both other cases.  This strongly suggests
that episodes of high accretion are stochastic.  As we show in
\S\ref{sec.clump_acc}, high-accretion rate episodes occur when the BH
collides with dense clouds of gas in the ISM.

In simulations with low-$f_{\rm gas}$, the accretion rate again hovers
around $10^{-5}$ Eddington, yet almost never jumps above the Eddington
limit.  Black hole growth is tiny and relatively constant, unlike in
the gas-rich galaxies where most BH growth occurs in brief episodes.
These low-$f_{\rm gas}$ simulations have not developed the stellar
bars that are relatively common in $z<1$ galaxies \citep[and rare at
  higher redshifts][]{sheth08, kraljic12}, which some authors argue
can drive significant gas inflows \citep[e.g][]{athanassoula92}.
Evidence for a link between bars and AGNs is mixed \citep{shlosman00,
  knapen00,lee12}, possibly because orbital resonances block the
inflow if a bulge and/or black hole is present
\citep[e.g.][]{combes85}.  We have performed a preliminary test
  of bar influence by running simulation M4f10 at low resolution until
  a stellar bar naturally develops (after about 1.5 Gyrs).  Once we
  run at high-resolution with the bar, we find that the average black hole
  accretion rate is only $\sim 60$\% higher than in our fiducial run,
  and therefore black hole growth remains essentially negligible.  The
  presence of a bar does not appear to change our results.

Thus, a dichotomy of BH growth emerges between high-$f_{\rm gas}$
(i.e. high-$z$) and low-$f_{\rm gas}$ (low-$z$) disks.  Gas-rich disks
occasionally feed rapidly-growing BHs, while BHs in gas-poor disks
remain quiescent \citep{dubois12_dual}.  We argue below that this dichotomy emerges due to
differences in ISM dynamics.  Gas-rich galaxies form dense, massive
clouds which can feed the black hole, while low-$f_{\rm gas}$ galaxies
do not.  Quantitatively, a $\sim5\times$ higher gas fraction leads to
$\sim10^3\times$ as much fractional BH growth.  All our simulations
show significant variability, a subject to which we return in \S
\ref{sec.feedback_impact}.

%
%
\subsection{BH accretion driven by collisions with dense clouds}
\label{sec.clump_acc}
\begin{figure*}
\includegraphics[width=168mm]{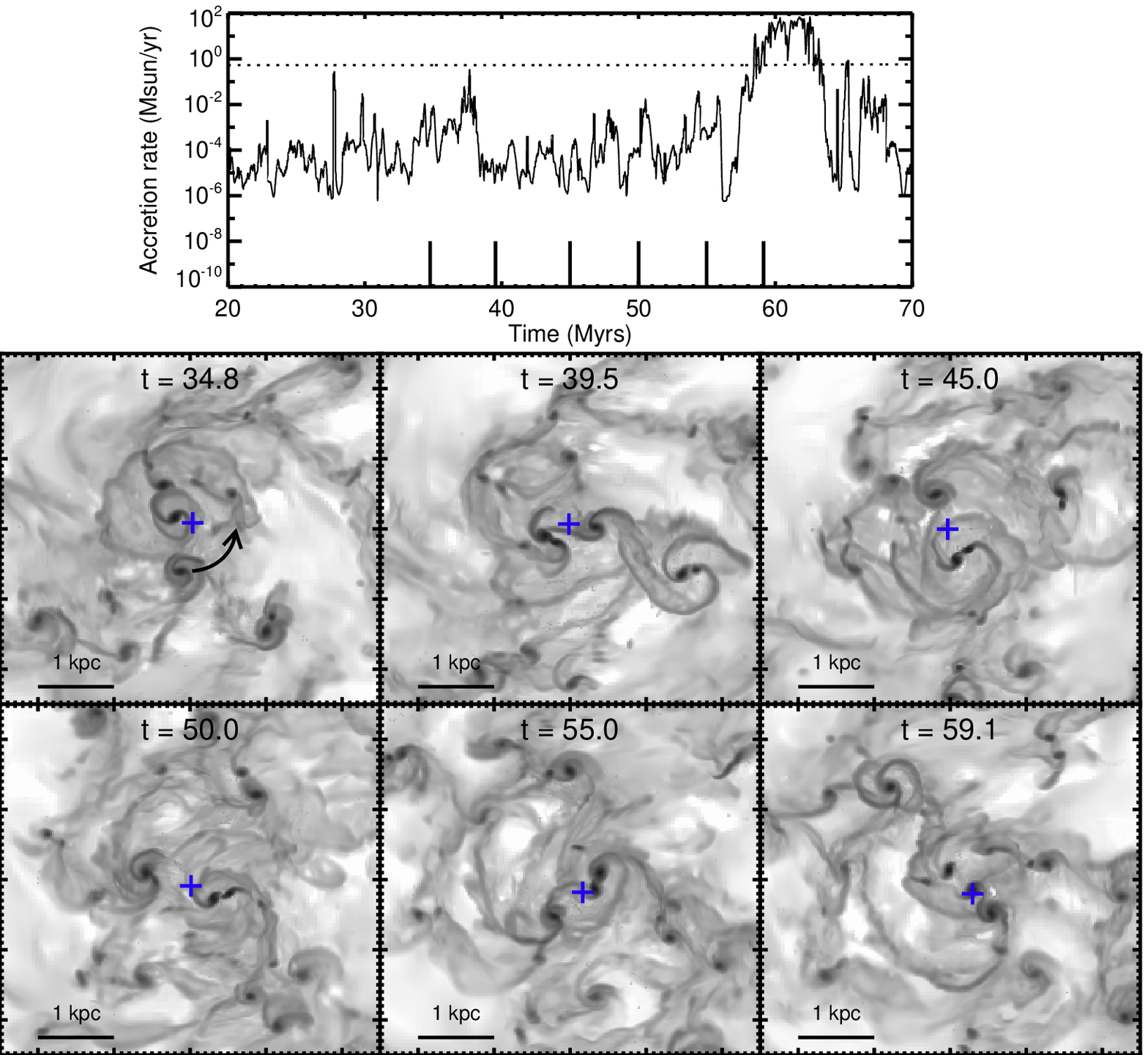}
\caption{ Collisions between the black hole and dense clouds in the
  ISM trigger episodes of Eddington-limited growth.  {\bf Top panel:}
  Bondi accretion rate as a function of time for simulation M4f50.
  {\bf Lower panels:} Images of gas density from six snapshots of the
  simulation separated by $\sim 5$~Myrs.  A blue cross indicates the
  location of the BH particle.  The time of each snapshot is shown at
  the top of each panel, and is marked as a vertical line on the time
  axis in the in the accretion rate panel above.  In the final
  snapshot (lower right, $t=59$~Myrs), a dense cloud has collided with
  the BH, driving the accretion rate to the Eddington limit.}
\label{fig.snapshots_plus_acchist}
\end{figure*}

\begin{figure*}
\includegraphics[width=168mm]{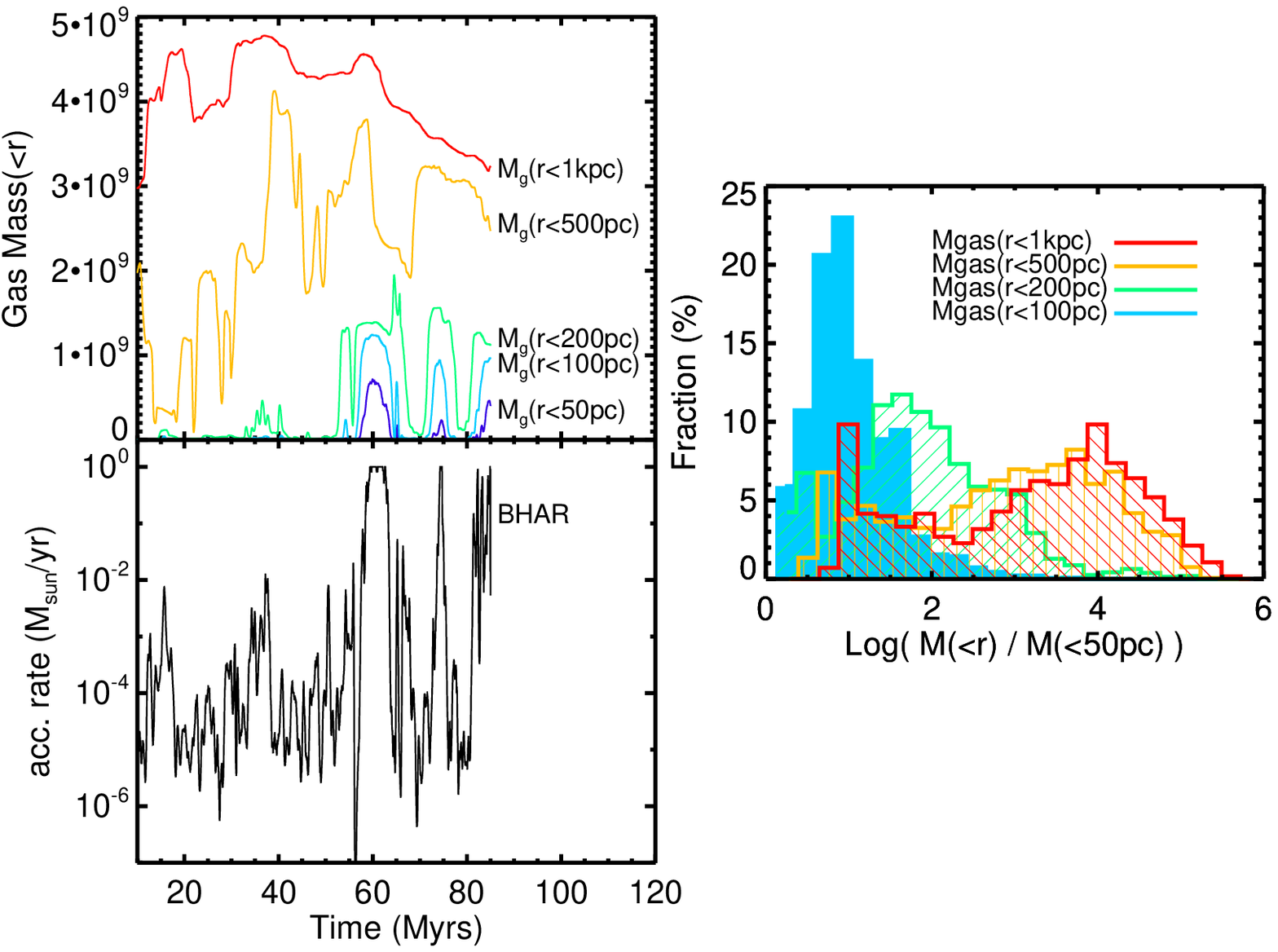}
\caption{ Black hole accretion traces gas inflow within 200pc, but is
  only loosely correlated with inflows within 1 kpc.  {\bf Top:} Total
  gas mass within radius $r$ of the BH in simulation M4f50, for five
  values of $r$, as labelled.  {\bf Bottom:} BH accretion rate as a
  function of time.  The BH accretion rate peaks when the gas mass
  within 200~pc peaks.  {\bf Right:} The distribution of $M_{\rm
    gas}(<r)/M_{\rm gas}(<50$~pc$)$, for four values of $r$.  The gas
  mass within 100~pc (blue histogram) is typically 10 times that
  within 50~pc.  The gas mass within 1~kpc is typically $10^4$ times
  that within 50~pc, with a broad distribution indicating that
  the correlation between scales is poor.}
\label{fig.gasmass_acc_all}
\end{figure*}

\citet{bournaud11} show that gas-rich disk galaxies undergo inflow
that can feed the central black hole, and argue that this inflow
occurs in both clumps and smooth flows.  In our simulations, periods
of high accretion rates are directly tied to dense structures within
the galaxy's ISM \citep[cf. ][]{dubois13}.

Figure \ref{fig.snapshots_plus_acchist} shows the infall of a dense,
massive cloud, and its effects on the BH accretion rate, in galaxy
M4f50.  The top panel shows the (smoothed) BH accretion rate as a
function time: this is a sub-interval of the accretion rate in Figure
\ref{fig.time_series}.  The lower six panels show the turbulent
galactic center at six different times (spaced by about 5~Myrs,
labelled), with a blue cross for the location of the BH.  The times
corresponding to the snapshots are marked as vertical lines along the
x-axis of the accretion rate panel above.

As multiple star-forming clouds rotate around the galactic center,
mutual interactions among the clouds cause them to scatter off one
another (the stellar galactic center, in which the BH sits, is also
scattered somewhat).  Filamentary structures between dense clouds
occasionally collide with the black hole, inducing fluctuations in the
accretion rate by 4 orders of magnitude, up to $>1$\% of the Eddington
limit.  Finally, one of the dense clouds collides with the black hole
($t\sim59$~Myrs), triggering a persistent phase of Eddington-limited
accretion.  Note that this dense cloud is \emph{not} one of the more
massive clumps in the simulation (seen in the outer disk in Figure
\ref{fig.example_snapshots}), and it is smaller and less massive
than the ``giant clumps'' identified in high-redshift observations
\citep{elmegreen05, elmegreen09, forster-schreiber11, guo12}.

At any given time, multiple dense clouds are within 1~kpc of the BH.
Yet Eddington-limited BH growth only occurs when such a cloud directly
collides with the BH.  This implies that BH accretion is only
indirectly linked to gas inflow at 1~kpc, and directly linked only to
the gas much closer in \citep{hopkins10_accretion, bournaud11}.
Figure \ref{fig.gasmass_acc_all} illustrates the relation between gas
mass at various scales and the BH accretion rate (BHAR).  The upper
panel shows the total gas mass within radius $r$ of the BH, for radius
values $r=50,100,200,500,$ and $1000$~pc, as a function of time.  Note
the masses ($y$-axis) are shown on a linear scale.  In the bottom
panel, we show the BH accretion rate on a log scale.  Within
$r=200$~pc, rapid increases in gas mass -- sometimes by factors of
$\sim 10$ -- clearly trigger increases in the accretion rate.  At
larger radii ($r=500$ or 1000 pc), the total gas mass changes more
slowly, and the connection between changes in the gas mass and the
accretion rate is weak.

In the right panel of Figure \ref{fig.gasmass_acc_all}, we show the
distribution of values of $M_{\rm gas}(<r) / M_{\rm gas}(<50$~pc$)$
for each $r>50$~pc.  This illustrates the connection between gas at
large scales and gas in the immediate vicinity of the BH.  The
connection is relatively strong for gas at 100~pc: there is typically
$10\times$ as much gas within $r=100$~pc as at $r=50$~pc, with a scatter
of a factor of 3.  At larger radii the connection is weaker, with the
distributions becoming broader.  The mass of gas within 1~kpc can be
as much as $10^5 \times$ the mass within 50~pc, and the scatter in the
distribution is a factor of 20.  Therefore, cosmological simulations
with a typical resolution of $\sim1$~kpc cannot estimate the gas mass
close to the BH within a factor of $\sim$20.  This poses problems for
typical implementations of BH fueling in such simulations
\citep[as argued by][]{hopkins10_accretion}.

The critical radius at which the BH accretion rate responds directly
to the local gas mass appears to be $\sim200$~pc in simulation M4f50.
This critical radius is approximately the characteristic size of dense
clouds seen in Figure \ref{fig.snapshots_plus_acchist}.  If they are
revolving about the galactic center at the galaxy's rotation velocity,
clouds outside this radius still must lose a huge fraction of their
angular momentum to feed the black hole.  Gas at 500~pc must lose
$\approx97$\% of its angular momentum to migrate within 100~pc; for
gas at 200~pc, this number is $\approx75$\%.  Dense clouds lose this
angular momentum via scattering with other clouds.  Due to
resolution limits we cannot address angular momentum loss within
$r_{\rm acc}\approx 24$~pc.

During the accretion event at $t\approx 60$~Myrs driven by collision
with a dense, massive cloud, the BH gains a total mass of $\sim2\times10^6
M_{\sun}$.  The gas mass of the dense cloud is $\sim10^9$, and the
mass within $\sim50$~pc is about half that.  Even when a cloud
collision triggers Eddington-limited feedback, the BH only accretes
$\sim1/1000$ of the cloud mass.

\subsubsection{A new mechanism for BH accretion?}

BH accretion driven by random collisions with dense clouds is similar
to the stochastic accretion described in \citet{hopkins_hernquist06},
but some of their assumptions about ISM structure and dynamics do not
hold in our simulations.  In their analytic model, small interstellar
clouds randomly collide with the BH, then AGN feedback immediately
enters the ``blowout'' phase which destroys the cloud.  In our
(high-$f_{\rm gas}$) simulations, a high accretion rate persists for
several Myrs rather than being immediately shut down by AGN feedback.
The simulated clouds colliding with the BH are sufficiently massive
and dense to avoid immediate disruption by AGN feedback (see
\S\ref{sec.feedback_impact}).  This enables the accretion rate to
remain near the Eddington limit for an extended period, and raises the
time-averaged accretion rate of the BH (as we show in
\S\ref{sec.implications}).  

The prominence of dense clouds in feeding the BH suggests that high
accretion rates rely on cloud survival.  If strong stellar feedback
were to disrupt the dense clouds before colliding with the black hole,
it is not clear that the BH would ever reach Eddington-limited
accretion rates.  We have mitigated this uncertainty by restricting
the duration of our simulations to less than the lifetimes of giant
clumps \citep{wuyts12, dekel13}.  The dense clouds responsible for most BH growth in
our simulations, however, are smaller and less massive than the observed ``giant''
clumps, and lifetimes of these more moderate clouds remain uncertain.
In future work we will test the impact of variations in stellar
feedback on the black hole accretion rates seen in these simulations.

Based on this dense cloud-driven accretion process, we predict that
AGNs are more common in unstable disks \citep{bournaud12}.  We do not
necessarily predict a one-to-one correlation between AGNs and the
presence of giant clumps in the outer disk, but rather with more
moderate dense clouds near the galactic center.  In our
  simulations, powerful AGN occur only when the BH is embedded within
  a dense cloud.  In the real Universe, such clouds would
  have substructure that determines whether the AGN is actually
  triggered, but our simulations suggest that a large reservoir of
  fuel provided by the dense cloud is required to maintain a powerful AGN.
We also expect that the dense clouds driving accretion will partly
obscure the AGN while they remain intact.  With up to $\sim 10^9
M_{\sun}$ of gas within 200~pc of the BH, we estimate column densities
$\sim 10^{23-24}$~H~cm$^{-2}$, enough to significantly attenuate the
emitted X-rays.  How this varies with line-of-sight is left to future
work.

In summary, massive, dense clouds drive BH growth in gas-rich
galaxies.  These clouds form as instabilities in gas-rich disks, and
through mutual interactions they migrate inwards as part of a global
inflow \citep{bournaud11}.  The dense clouds stochastically collide
with the BH, inducing persistent phases of Eddington-limited BH
growth.

%
%
\subsection{Impact of AGN feedback on accretion rates}
\label{sec.feedback_impact}

\begin{figure*}
\includegraphics{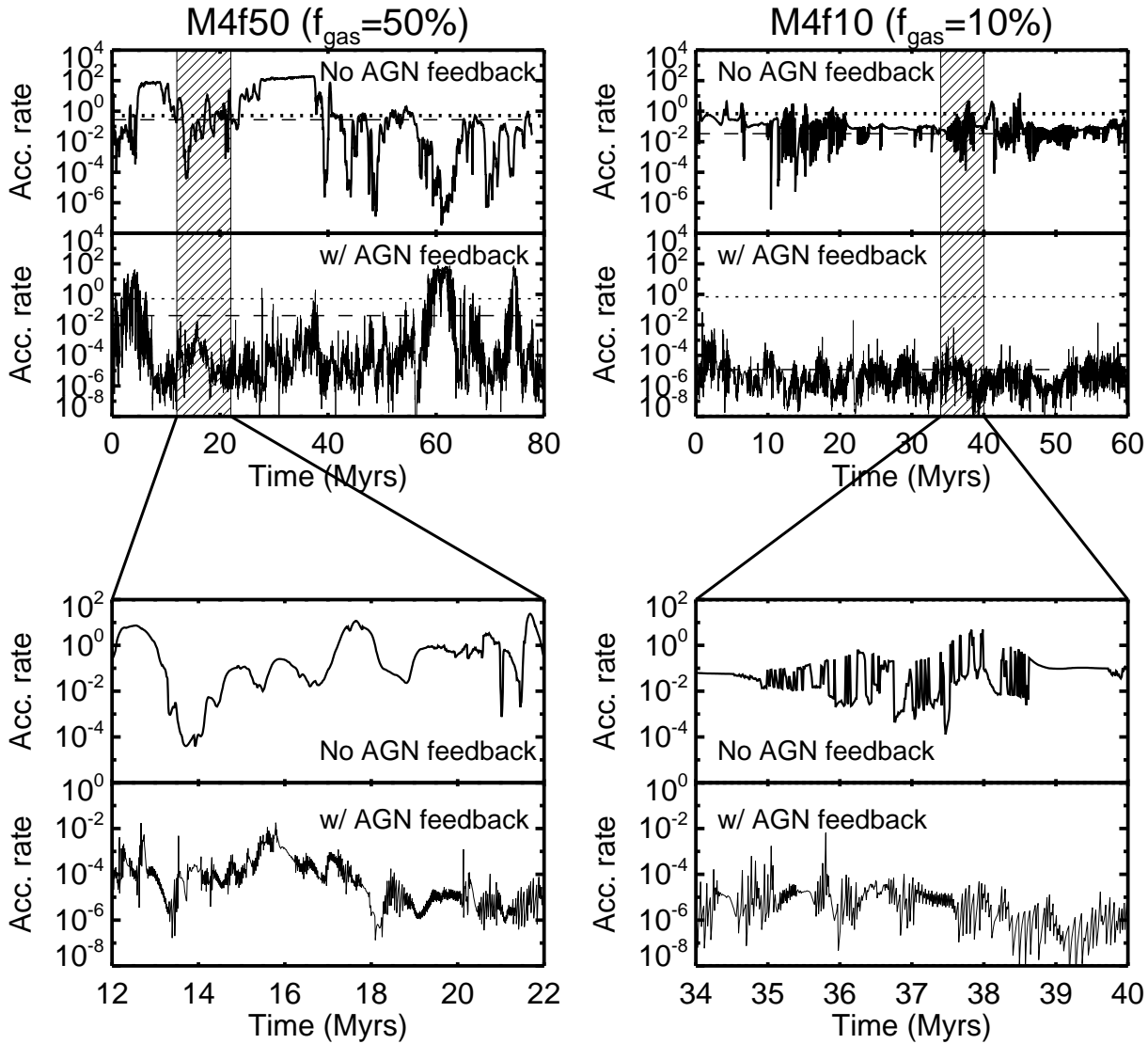}
\caption{The Bondi accretion rate (in $M_{\sun}$~yr$^{-1}$) as a
  function of time for simulations M4f50 (high-$f_{\rm gas}$, left
  column) and M4f10 (low-$f_{\rm gas}$, right column), without (top
  panels) and with AGN feedback (the fiducial case).  The bottom
  panels show a zoom-in of accretion rates in both simulations to
  clarify the variability.  AGN feedback drastically reduces average
  accretion rates, but accretion rate variability on $\sim Myr$
  timescales is driven mainly by the clumpy structure of the ISM.  In
  the low-$f_{\rm gas}$ simulation (right column), the presence of AGN
  feedback prevents phases of Eddington-limited accretion that occur
  in the ``no feedback'' case. }
\label{fig.compare_acc_agn_noagn}
\end{figure*}

\begin{figure}
\includegraphics{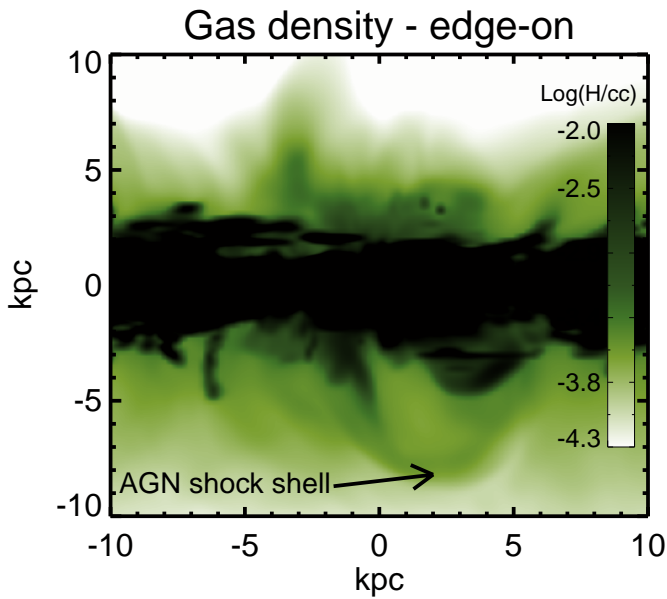}
\caption{An edge-on image of simulation M4f50 shows AGN-driven
  outflows $5-10$~kpc away from the disk plane.  This image is from
  time $t\approx70$ Myrs, shortly after a burst of Eddington-limited
  accretion.  The AGN drives a high-velocity outflow which shocks the
  surrounding gas, forming a shell below the disk.  We optimize the
  image scale (the stretch) to highlight the low-density outflow features, but this
  causes the dense gas near the disk to appear saturated.}
\label{fig.outflow}
\end{figure}

As in other simulations with AGN
\citep[e.g.][]{dimatteo05,springel05, degraf10, booth09, debuhr10}, black
hole growth is self-regulated by AGN feedback.  In this section we
directly illustrate the effects of AGN feedback, and show that it
dramatically reduces accretion rates.  We also show that AGN feedback
imposes very short timescale variability in our simulations, but that
variability on $\sim$~Myr scales is driven by structure in the ISM.

Figure \ref{fig.compare_acc_agn_noagn} compares the BH accretion rates
in our fiducial simulations to simulations without AGN feedback.  The
simulation pairs shown here start from the same point, $t=0$.  During
the relaxation phase preceding this time (i.e. $t<0$), AGN feedback
is switched on in both cases.  After $t=0$, AGN feedback is switched
off in the ``No AGN feedback'' simulation.  Both simulations use an Eddington
limit on the accretion rates, and both black holes are allowed to grow
in mass.  The only difference is that in the ``No AGN feedback'' simulation, we
do not inject thermal energy associated with black hole growth.

As shown in Figure \ref{fig.compare_acc_agn_noagn}, the ``No AGN
feedback'' simulations have generally higher Bondi accretion rates,
with a larger fraction of the time spent above the Eddington limit.
Periods of high accretion in the ``No AGN feedback'' simulations do
\emph{not} correspond one-to-one to high-accretion periods in the fiducial
simulation.  There are some periods where the accretion rate
\emph{with} feedback is \emph{higher} than without feedback
(e.g. around $t=60$~Myrs).  This is apparently because AGN feedback
alters the formation and dynamics of dense gas clouds in the galactic
center.

In the ``No AGN feedback'' simulation of low-$f_{\rm gas}$ galaxy
M4f10, accretion rates are again significantly higher and even beyond
the Eddington limit.  This indicates that AGN feedback
prevents Eddington accretion in low-$f_{\rm gas}$ galaxies.  We
speculate that in low-$f_{\rm gas}$ galaxies, feedback disrupts the
(relatively small) clouds that otherwise would rapidly feed the black hole.
High-$f_{\rm gas}$ galaxies form more massive, denser clouds that the
feedback cannot disrupt, and these ultimately fuel Eddington
accretion.  We leave the details of cloud disruption by feedback to
future work.

In \S \ref{sec.clump_acc} we mentioned the high accretion rate
variability in our fiducial simulations.  Similar variability is seen
in other simulations with and without AGN feedback \citep{levine10,
  hopkins10_accretion, novak11}.  Some of the variability, especially
on the shortest timescales, is induced by our AGN feedback model.
Figure \ref{fig.compare_acc_agn_noagn} shows that even without AGN
feedback, the BH accretion rate remains highly variable, though on
longer timescales than in the fiducial simulations.  The accretion rate
sometimes changes by $\sim5$ orders of magnitude over $<1$~Myr.  This
implies that the accretion rate variability (on these timescales) is
driven by the clumpy and filamentary structure of the ISM rather than
AGN feedback.  The characteristic variability timescale is determined
by the typical size and velocity of a gas structure.  A typical dense cloud
of size $100$~pc with a velocity of $50$~\kms (which is typical for
central clouds in M4f50) will pass over the black hole in $\sim
2$~Myrs, which is similar to the variability timescale we see.  The
``no feedback'' M4f10 (low-$f_{\rm gas}$) simulation shows variability
on shorter timescales ($\sim 0.1$~Myr) owing to smaller typical sizes of ISM
structures.

AGN feedback introduces an additional high-frequency variability that
is a direct consequence of our feedback implementation.  The blast
region into which the AGN injects its energy is the same as the region
over which the accretion rate is measured.  Thus, by heating up the
gas we directly lower the Bondi rate.  The AGN blast is only executed
at each coarse timestep, which is the timescale on which our feedback
model induces variability in accretion rates (technically, the coarse timestep
varies based on conditions in the simulation).  This high-frequency
variability depends on the details of the feedback implementation, and
thus may not be physically accurate.

The AGN feedback does not have a \emph{major} impact on the
star-forming gas in the galactic disk \citep[except perhaps in the central
regions; cf.][]{newton13}, but it does drive high-velocity outflows from the galaxy.
We show an example outflow in Figure \ref{fig.outflow}.  We show an
edge-on view of simulation M4f50, with the image stretch adjusted to
highlight low-density structures outside the disk.  The
Eddington-limited accretion event at $t=60$~Myrs has driven an outflow
that shocks the gas below the plane of the disk.  This AGN-driven
shock shell has expanded $\sim 8$~kpc below the disk in $\sim10$~Myrs,
implying an average outflow velocity of $\sim750$~\kms.  Above the
disk, gas densities are lower (due to earlier feedback
events), allowing chimneys of gas to escape.  In future work we will
study these outflows, and the broader impact of AGN feedback on
galaxies, in detail.

In summary, AGN feedback suppresses BH accretion and imposes some
variability, but ISM structure drives accretion variability on
$\sim1$~Myr timescales.  The AGN feedback \emph{cannot} disrupt the
dense, massive clouds in gas-rich galaxies, so collisions with these
clouds trigger persistent Eddington accretion.  AGN feedback
\emph{can} disrupt the smaller gas clouds in low-$f_{\rm gas}$
galaxies, thereby preventing high accretion rates.  Although AGN feedback does
not have a drastic effect on the star-forming gas in the outer disk,
it does drive high-velocity outflows.

\section{Implications for BH-galaxy co-evolution}
\label{sec.implications}
Although our suite of simulations is small, our broad results can be
extended to shed light on the nature of black hole fueling in the
broad context of galaxy evolution.  One crucial result we have already
discussed is that black holes grow significantly in isolated, gas-rich
disks.  In this section, we examine Eddington ratio distributions in
our simulations, quantify the implied AGN duty cycle, and study the
relationship between SFR and accretion rate.

%
%
\subsection{Eddington ratio distributions}
\begin{figure}
\includegraphics[width=80mm]{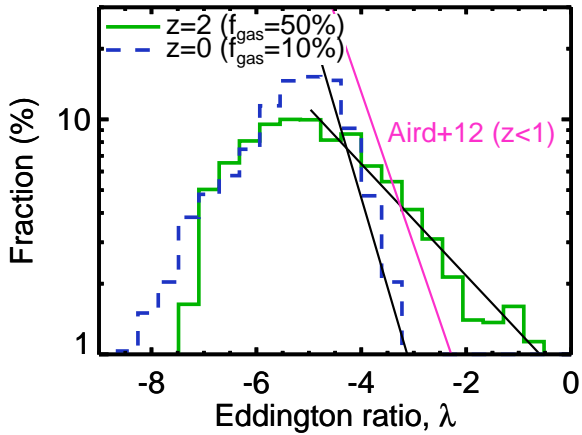}
\caption{Eddington ratio distributions averaged for gas-rich $z=2$
  (solid green) and gas-poor $z=0$ (dashed blue) simulations.  The
  distributions take a roughly log-normal shape.  We fit a power-law
  to the high-$\lambda$ tail of each distribution (thin solid lines),
  and compare with the observed distribution for $z\lesssim1$ galaxies
  from \citet[][pink line]{aird12}, including a normalization
  correction (see text).  The observed distribution has a similar
  slope to our low-$f_{\rm gas}$ $z=0$ galaxies, but generally higher
  accretion rates probably owing to galaxies with $f_{\rm gas}>10$\%.}
\label{fig.edd_dist}
\end{figure}

The Eddington ratio quantifies BH accretion activity independent of BH
mass, and its accessibility to observational estimates makes it a
useful tool to understand the life cycles of BH accretion.  Observers
estimate the Eddington ratio $\lambda=L_{\rm AGN,bol}/L_{\rm edd}$
using either the full spectral energy distribution \citep{trump11} or
a bolometric correction \citep{heckman04, vasudevan07, kauffmann09,
  aird12} for the AGN luminosity $L_{\rm AGN,bol}$, and a BH mass
measurement for the Eddington limit.  BH mass estimates typically come
from assuming scaling relations between galactic bulges and BH mass,
but for quasars better estimates may be available from emission line
scaling relations \citep[e.g.][]{kelly07}.  In our simulations the
Eddington ratio is just the actual accretion rate divided by
the Eddington rate: $\lambda = \dot{M}_{\rm acc} / \dot{M}_{\rm Edd}$.

Figure \ref{fig.edd_dist} shows Eddington ratio distributions of our
simulations.  Episodes of high accretion are rare and stochastic, and
the $100$~Myr duration of each simulation does not adequately sample
these events -- simulation M16f50 has only one high-accretion episode,
whereas M4f50 has several.  In order to correct for this
stochasticity, we take the combined average Eddington distribution for
the high-$f_{\rm gas}$ galaxies, and a separate average for the
low-$f_{\rm gas}$ galaxies.  This is what we show in Figure
\ref{fig.edd_dist}.

In observations, Eddington ratios are difficult to measure at very low
values ($\lesssim 10^{-5}$), so we follow observers in fitting a
power-law to the high-$\lambda$ end of each distribution.  The power-laws (straight solid
lines) provide good fits with a slope of $-0.24$ for the $f_{\rm
  gas}=50$\% simulations, and a slope of $-0.76$ for the $f_{\rm
  gas}=10$\% simulations.  As expected, the high-$f_{\rm gas}$
galaxies are much more likely to have Eddington ratios $>0.01$ than
are the low-$f_{\rm gas}$ galaxies.

We compare with the power-law observational Eddington ratio
distribution of \citet{aird12} for galaxies with $0.2<z<1.0$.  We use their best-fit slope of
$-0.65$ \citep[which is consistent with that from][]{kauffmann09}, and
make a rough correction so that the normalization is comparable to our
low-$z$ distribution.  In our low-$f_{\rm gas}$ distribution, 46\% of the
time is spent at Eddington ratios $>10.0^{-5}$, while in the
observations all AGN are above that level.  Thus we multiply the
best-fit normalization from \citet{aird12} by $1.0/0.46$ as a
normalization correction.  The result is shown as a pink line in
Figure \ref{fig.edd_dist}.  The observed slope is quite similar to
that of our low-$f_{\rm gas}$ simulations, but the normalization falls
between our high- and low-$f_{\rm gas}$ curves.  

With redshifts up to $z=1$, the observed sample probably includes many
galaxies with $f_{\rm gas}>10$\% (as well as mergers) that could
explain why the observations show higher accretion rates than our
simulations. \citet{aird12} do note an increase in the fraction of
galaxies with high Eddington ratios at higher redshifts, although they
suggest the same power-law slope holds at all observed redshifts.  Our
simulations predict that, along with the increased fraction of
high-accretion events, the slope evolves with gas fraction
(i.e. redshift), becoming significantly flatter for galaxies at
$z\sim2$ than for low-redshift galaxies.

%
%
\subsection{The AGN duty cycle and black hole growth}
\label{sec.duty_cycle}
\begin{figure*}
\includegraphics[width=168mm]{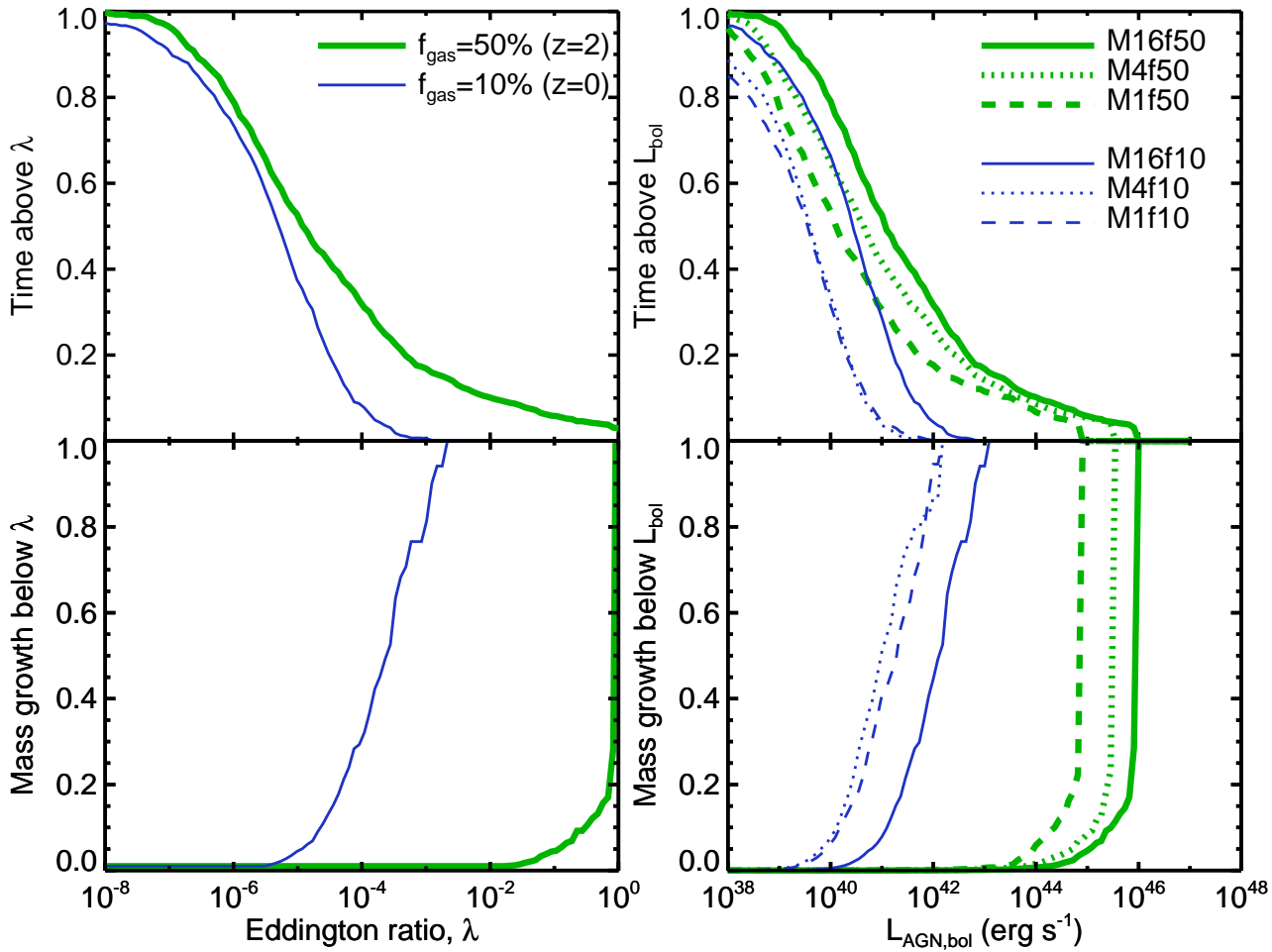}
\caption{BHs spend a small fraction of their time at high accretion
  rates, but mass growth is dominated by these high-accretion phases.
  {\bf Top row:} Fractional time spent above Eddington ratio $\lambda
  = \dot{M}_{\rm acc} / \dot{M}_{\rm Edd}$ as a function of Eddington
  ratio (left column), and time spent above bolometric AGN luminosity
  $L_{\rm bol}$ (right column).  Lines in the left panels come from
  the average Eddington ratio distributions in Figure
  \ref{fig.edd_dist}.  In the right panels, for each simulation (lines
  as labelled), we show the AGN luminosities assuming that each BH
  draws from the \emph{average} Eddington ratio distribution for its gas
  fraction (see text).  {\bf Bottom row:} Fractional black hole mass
  growth that occurs below a given Eddington ratio (left column) and
  below a given bolometric luminosity. }
\label{fig.time_above}
\end{figure*}

The AGN duty cycle plays a central role in efforts to understand AGN
triggering (and thus black hole fueling).  The duty cycle can be
loosely thought of as the fraction of galaxies whose AGN is ``on'' at
a given time, or the fraction of time a given galaxy's AGN is ``on.''
Observationally, an AGN appears to be ``on'' only if it is
sufficiently bright to be detected in whatever wavelength is being
observed (e.g. X-rays).  Estimates of the duty cycle vary with
selection technique, luminosity/accretion rate thresholds, redshift,
and BH mass \citep[cf.][]{marconi04, shankar09}.

Simulations (including ours) indicate that AGNs have complex light curves,
so the duty cycle is most conveniently defined by quantifying the
fraction of time spent above a given luminosity or accretion rate
\citep[e.g.][]{hopkins05_lifetimes}.  Using the Eddington ratio
distributions shown in Figure \ref{fig.edd_dist} (plus knowledge of
the BH masses from the simulations), we calculate the duty cycle in
this way.

In the top row of Figure \ref{fig.time_above} we show the fraction of
time spent above a given Eddington ratio (left panel) and AGN
bolometric luminosity (right panel).  The Eddington ratios, averaged
over galaxies of a given gas fraction, are taken directly from Figure
\ref{fig.edd_dist}; Figure \ref{fig.time_above} essentially presents
the same data in a cumulative way.  

We calculate AGN luminosities as
follows.  We apply the combined average Eddington ratio distribution
for gas-rich galaxies (green curve in Figure \ref{fig.edd_dist}) to
each of the gas-rich galaxies for a hypothetical 100 Myr period.
Starting from the original $t=0$ BH masses from Table
\ref{table.runs}, we advance through 1000 steps of 0.1 Myrs.  At each
step we randomly select an Eddington ratio from the average
distribution and apply the corresponding accretion to the BH growth.
We calculate the AGN luminosity from the accretion rate via $L_{\rm
  AGN,bol} = \epsilon_r \dot{M}_{\rm acc} c^2$, where we use
$\epsilon_r=0.1$.  This method corrects for stochasticity by assuming
that each individual galaxy takes on the combined average Eddington
ratio distribution of galaxies with similar gas fractions  \citep[see also][]{hickox13}.

From Figure \ref{fig.time_above} (as well as Figure
\ref{fig.edd_dist}) we see that black holes in gas-rich galaxies
(thick green lines) typically spend more time at higher Eddington
ratios than their low-$f_{\rm gas}$ counterparts (thin blue lines).
Furthermore, the gas-rich galaxies spend $20-40$\% of the time above
$L_{\rm AGN,bol}=10^{42}$~erg~s$^{-1}$, and $\sim10$\% of the time
above $10^{44}$.  A correction from bolometric luminosity to X-ray
luminosity of $L_{\rm bol} \approx30 L_{X}$
\citep[cf.][]{barger01,marconi04} implies that our gas rich galaxies
should spend $10-15$\% of the time detectable as X-ray AGNs (with
$L_{X}> 10^{42}$~erg~s$^{-1}$), and up to 25\% of the time as weak (or
stronger) AGNs with $L_{X}> 10^{41}$~erg~s$^{-1}$.  Our low-$f_{\rm
  gas}$ galaxies spend less than 1\% of the time detectable as X-ray AGNs.

In the bottom row of Figure \ref{fig.time_above}, we show the fraction
of the total BH mass growth attributed to accretion below a given
value of the Eddington ratio (left) or the AGN luminosity (right).

In gas-rich simulations (thick green lines), nearly all of the BH mass
growth occurs at Eddington ratios above $10^{-2}$, and 80\% of the growth
occurs at Eddington ratios $\gtrsim 0.95$.  In low-$f_{\rm gas}$
simulations, on the other hand, the (meagre) BH growth occurs at
Eddington ratios near $10^{-4}$.  The factor of $\sim5$ difference in
gas fraction leads to a difference in Eddington ratios of up to
$10^4$.  This large gap reinforces the dichotomy of BH growth between
high- and low-$f_{\rm gas}$ galaxies.  BH fueling in gas-rich galaxies
is dominated by a few dense clouds that migrate to the center, whereas
fueling in gas-poor galaxies is much smoother.

Although we have presented a small number of simulations, these duty
cycle data can be incorporated into semi-analytic models of BH-galaxy
co-evolution \citep[e.g.][]{kauffmann00, volonteri03, somerville08,
  hirschmann12}.  The Eddington distributions provide a quantitative
estimate of the black hole growth that occurs in isolated disks.  We
hope to construct larger suites of simulations, encompassing a better
sampling of galaxy properties, in the future.

%
%
\subsection{Coincident AGN and star-formation}
\begin{figure}
\includegraphics[width=80mm]{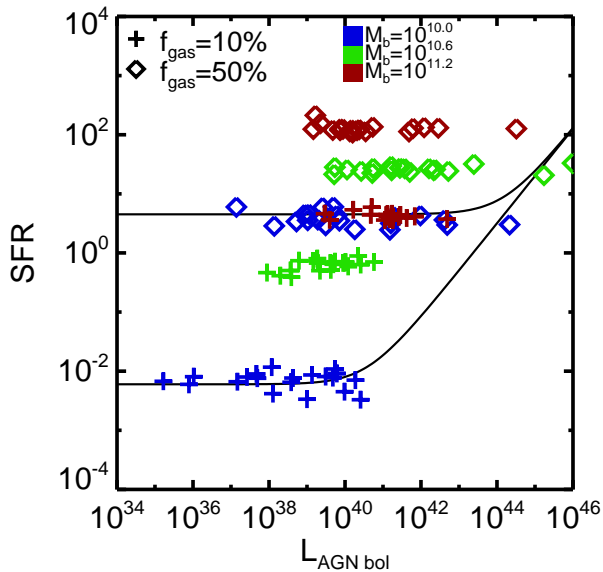}
\caption{ SFR versus AGN luminosity at 20 snapshots from each of our
  six simulations.  High-$f_{\rm gas}$ galaxies (diamonds) have higher
  SFRs and reach higher AGN luminosities than their low-$f_{\rm gas}$
  counterparts (crosses).  Points are colour-coded by baryonic mass as
  shown in the legend.  Solid lines show the SFR$-L_{\rm AGN}$ trend
  (horizontal line) and envelope (log slope of 0.8) for our
  lowest-mass galaxies at both high and low redshift, following
  \citet{rosario12}.}
\label{fig.sfr_vs_LAGN}
\end{figure}

Recent observations link AGN to star-formation \citep[e.g.][]{rosario12,
  mullaney12}.  The observations suggest a scenario where Main Sequence
star-forming galaxies occasionally fuel their black holes without a
trigger that substantially alters their star-formation rates.  We have
already demonstrated that high-redshift star-forming disk galaxies can
host AGNs without any external triggers such as mergers.  In this
section, we connect directly to observable star-formation rates and
AGN luminosities.

Figure \ref{fig.sfr_vs_LAGN} shows the SFR as a function of bolometric
AGN luminosity at 20 evenly-spaced times from each of our simulations.
In an individual simulated galaxy, the SFR changes only slightly, but the AGN
luminosity varies by $\sim6$ orders of magnitude within a single
dynamical time.  As observed for X-ray selected AGN by
\citet{rosario12}, AGN at low luminosities may live in galaxies with a
wide spread in SFR.  At high AGN luminosities there is an envelope
such that high-luminosity AGN live only in galaxies with high-SFR.  We
illustrate this trend in the figure with solid lines of the form ${\rm
  SFR}=A + B(L_{\rm AGN}^{0.8})$, where $A$ and $B$ are arbitrary
constants chosen to match our low-mass galaxies, and the power law slope
of 0.8 is approximately the best-fit value from \citet{rosario12}.

Two effects combine to form this envelope: a mass effect, and a gas
fraction effect.  First, galaxies with higher stellar masses host more
massive black holes, and these more massive black holes accrete gas at
higher absolute rates (for a given Eddington ratio) than lower-mass
black holes.  This effect may be sensitive to the accretion model --
the Bondi formula depends strongly on black hole mass, but not all
accretion models do \citep{hopkins10_accretion}.  Second, black holes in
high-redshift galaxies with higher gas fractions spend more time at
high Eddington ratios.  This results from the dichotomy in ISM
properties that depends on gas fraction -- gas rich galaxies form
dense clouds that fuel Eddington-limited accretion.

%
%
\subsection{Cosmological growth of BHs in disk galaxies}
\label{sec.cosmo_growth}
\begin{figure}
\includegraphics[]{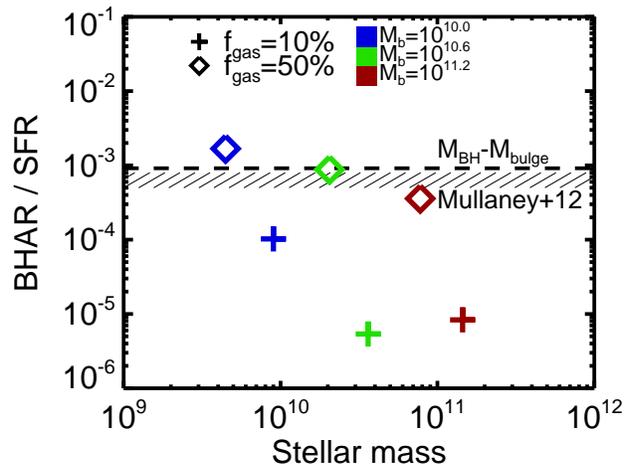}
\caption{Relative growth rate of black hole mass to stellar mass as a
  function of stellar mass for our simulations (crosses: low-$f_{\rm
    gas}$, diamonds: high-$f_{\rm gas}$).  The $y-$axis is
  $\dot{M}_{\rm BH} / \dot{M}_{\rm stellar}$, or the BH accretion rate
  divided by the SFR, where each is averaged over the duration of the
  simulation.  In order to correct for stochasticity of accretion
  events, the BH accretion rates assume that each BH draws from the
  average Eddington ratio distribution for its gas fraction (see
  text).  A horizontal dashed line represents the growth rates
  required for the galaxy to end up on the observed local $M_{\rm
    BH}-M_{\rm bulge}$ relation.  The hatched area corresponds to the
  observed range from \citet{mullaney12} for $z=1-2$ galaxies, which
  is comparable to $z=0$ results \citep{heckman04}.  Gas-rich ($z=2$)
  galaxies undergo sufficient BH growth to stay on the $M_{\rm
    BH}-M_{\rm bulge}$ relation, but low-$f_{\rm gas}$ galaxies show
  only $\sim1-10$\% of the required BH growth.}
\label{fig.relative_growth}
\end{figure}

While we have demonstrated that BHs can undergo significant,
Eddington-limited growth in isolated gas-rich disk galaxies, in this
section we address the role of this mode of BH growth in growing the
total mass of BHs in the present Universe.  Given the black hole mass
from the local $M_{\rm BH}-M_{\rm bulge}$ relation, what fraction of
the BH growth occurred in isolated disk galaxies?  A more detailed
exploration of the AGN luminosity function and the BH mass function
awaits future work with semi-analytic models.  Here we follow
observers \citep{heckman04, silverman09, mullaney12} in comparing the
BH growth rates ($\dot{M}_{\rm BH}$, i.e. the accretion rate) to the
stellar mass growth rates ($\dot{M}_{\rm stellar}$, i.e. the SFR).

Galaxies on the star-forming Main Sequence at $z=2$ (like our gas-rich
simulations) are likely to end up as red elliptical galaxies at $z=0$.
If they were to remain on the Main Sequence, they would generally grow
too much in stellar mass to explain the local galaxy stellar mass
function \citep{peng10}.  Therefore most of their ongoing
star-formation at $z=2$ contributes to the ``bulge'' mass -- i.e. the
total mass of an elliptical galaxy -- at $z=0$. (For this simple
analysis, we ignore any distinction between true bulges and
pseudobulges, with which BHs do not correlate; see e.g. \citealt{kormendy04, kormendy11}).  In order to satisfy the
$z=0$ $M_{\rm BH}-M_{\rm bulge}$ relation, the BH must eventually
undergo an equivalent amount of growth.  During the simulations, the
bulge growth (and BH growth) are small, ranging from insignificant to
$\sim 20$\% -- too small for a galaxy to move off the relation.  Thus,
we examine the average growth rates rather than the total growth in BH
mass and stellar mass.

Figure \ref{fig.relative_growth} shows $\dot{M}_{\rm BH} /
\dot{M}_{\rm stellar}$ as a function of stellar mass for our 6
simulations.  The BH accretion rates include a correction for
stochasticity by drawing their Eddington ratios from the average
Eddington ratio distributions in Figure \ref{fig.edd_dist} (see
\S\ref{sec.duty_cycle}).

For comparison, we also show the rough band of $\dot{M}_{\rm BH} /
\dot{M}_{\rm stellar}$ observed for $z=1-2$ galaxies
\citep{mullaney12}.  The local $M_{\rm BH}-M_{\rm bulge}$ relation
implies that the bulge is about $10^3\times$ as massive as the BH it
hosts -- \citet{bennert11} put this number at about 2200, with a
substantial uncertainty \citep{graham12}.  Assuming the black hole and stellar mass
grow in tandem, and accounting for stellar mass loss via stellar
evolution (winds and supernovae), we arrive at a rough expectation
that the time-averaged BH accretion rate must be $\sim 9\times10^{-4}$
of the SFR.  We plot this as a horizontal dashed line in Figure
\ref{fig.relative_growth}.  This can be thought of as the black hole
growth required to ``keep up'' with star-formation in order to end up
on the $M_{\rm BH}-M_{\rm bulge}$ relation at $z=0$.  As noted by
\citet{mullaney12}, their observations are close to the required BH
growth, implying that most BH mass is built-up in star-forming galaxies.

In Figure \ref{fig.relative_growth}, our gas-rich galaxies fall within
a factor of $\sim2$ of both the observed band and the required ratio
to keep galaxies on the $M_{\rm BH}-M_{\rm bulge}$ relation.  This
result implies that BH growth in isolated Main Sequence galaxies at
$z=2$ dominates BH growth at this epoch -- no mergers required.
Assuming that the \citet{mullaney12} observations indeed probe the
bulk of BH growth, BH fueling via collisions with dense clouds (as
described in \S\ref{sec.clump_acc}) could dominate the cosmic build-up of BH mass.


The low-$f_{\rm gas}$ galaxies show $\sim 1$\% of the required BH
growth.  This is not surprising since they never reach the
Eddington-limited accretion that dominates BH growth.  This result
suggests that BH accretion in isolated, low-redshift disk galaxies is
not a major contributor to BH growth, and that processes other than
disk instabilities \citep[e.g. mergers;][]{dimatteo05,
  springel05_mergers_ellipticals} must drive the observed BH growth in
these galaxies.  As noted previously, bar instabilities (which do not
occur in our fiducial simulations) could contribute to BH growth in
principle, but the evidence and our preliminary tests suggest that
bars are not major drivers of AGNs.

We must also note that the $M_{\rm BH}-M_{\rm bulge}$ relation may
vary with redshift.  Most observations suggest that, at high-$z$, BHs
are too massive to lie on the BH-bulge relations
\citep[e.g.][]{treu04,treu07,peng06,merloni10,bennert11_evolve}.
Ignoring the conflict this creates with our simulation initial
conditions, these observations imply that (at least some) black holes
should undergo a phase of growth that puts them above the $M_{\rm
  BH}-M_{\rm bulge}$ line in Figure \ref{fig.relative_growth} at
higher redshifts, then below the line at lower redshifts.

In summary, BHs in $z=2$, $f_{\rm gas}=50$\% simulations accrete sufficient
mass to keep them on the $M_{\rm BH}-M_{\rm bulge}$ relation and to
match observations of $\dot{M}_{\rm BH} / \dot{M}_{\rm stellar}$.  BHs
in $f_{\rm gas}=10$\% galaxies only grow $\sim1$\% of the required
amount, implying the need for other AGN triggers such as mergers.



\section{Summary and Conclusion} 
\label{sec.summary} 
We have simulated 6 isolated disk galaxies, with a range of baryonic
masses, at 6~pc resolution and a model for black hole accretion and
feedback.  Our simulations fall into two broad regimes: gas-rich
($f_{\rm gas}\approx50$\%) $z\sim2$ clumpy disks, and relatively
gas-poor ($f_{\rm gas}\approx10$\%) $z\sim0$ stable disks.  

Gas-rich $z\sim2$ disks form dense, massive clouds that migrate inward
due to mutual angular momentum exchange with other density
perturbations (Figure \ref{fig.snapshots_plus_acchist}).  The black
hole accretion rate fluctuates stochastically on $\sim1$~Myr
timescales as clouds and filaments collide with the BH (Figures
\ref{fig.time_series} and \ref{fig.compare_acc_agn_noagn}).
Occasionally a massive, dense cloud collides with the BH (Figure
\ref{fig.snapshots_plus_acchist}).  AGN feedback is insufficient to
disrupt these dense clouds, so the collision drives persistent ($\sim
10$~Myr) Eddington-limited accretion (Figure \ref{fig.time_series}).
The BH accretion rates correlate with gas mass fluctuations within
radii up to $\sim200$~pc, roughly the sizes of the clouds (Figure
\ref{fig.gasmass_acc_all}).  Owing largely to these dense cloud-driven
accretion events, BHs in gas-rich galaxies spend $\sim10$\% of their
time at Eddington ratios above $10^{-2}$ (Figures \ref{fig.time_above}
and \ref{fig.edd_dist}).  On average this growth agrees with $z=1-2$
observations, and is sufficient for the BH growth to ``keep up'' with
star-formation so that the galaxy ends up on the local $M_{\rm
  BH}-M_{\rm bulge}$ relation (Figures \ref{fig.sfr_vs_LAGN} and
\ref{fig.relative_growth}).  This dense-cloud driven accretion could
thus play a crucial role in the cosmic growth of BHs.

Low-$f_{\rm gas}$, low-redshift disks show markedly different
accretion patterns.  These galaxies develop spiral arms and small
clouds rather than massive clumps (Figure
\ref{fig.example_snapshots}).  These small density perturbations drive
stochasticity in the accretion rate on shorter timescales
($\sim0.1$~Myrs), and they are more easily disrupted by AGN feedback
(Figure \ref{fig.compare_acc_agn_noagn}).  The feedback effectively
eliminates episodes of Eddington-limited accretion, and these BHs
rarely even reach $10^{-2}\times$ Eddington (Figures
\ref{fig.time_above} and \ref{fig.edd_dist}).  With their growth so
suppressed, these BHs cannot ``keep up'' with their hosts' SFRs,
implying that additional AGN triggers such as mergers are required for
low-$f_{\rm gas}$ disks (Figure \ref{fig.relative_growth}).


 
 

\section*{Acknowledgements}
We thank Romain Teyssier for making RAMSES available, Florent Renaud
for help with code, and the anonymous referee for comments on the
paper.  We acknowledge support from the EC through grants
ERC-StG-257720 and the CosmoComp ITN.  Simulations were performed at TGCC
and as part of a GENCI project (grants 2011-042192 and 2012-042192).

\bibliographystyle{mn2e} 

\bibliography{paper}


\label{lastpage}

\end{document}